\documentstyle[12pt,aasms4]{article}

\lefthead{Preece et al.}
\righthead{BATSE High-Energy Spectroscopy}

\begin{document}

\title{BATSE Observations of Gamma-Ray Burst Spectra. IV.\\
Time-Resolved High-Energy Spectroscopy}

\author{Robert D. Preece, Geoffrey N. Pendleton, Michael S. Briggs, 
Robert S. Mallozzi, and William S. Paciesas}
\affil{Dept. of Physics, University of Alabama in Huntsville, 
Huntsville, AL  35899}

\author{David L. Band and James L. Matteson}
\affil{Center for Astrophysics and Space Sciences 0424, \\
University of California at San Diego,
La Jolla, CA 92093-0424}

\and

\author{C. A. Meegan}
\affil{NASA / Marshall Space Flight Center, ES84, Huntsville, AL  35812}


\begin{abstract}
We report on the temporal behavior of the high-energy power law continuum
component of gamma-ray burst spectra with data obtained by the Burst and
Transient Source Experiment. We have selected 126 high fluence and high flux
bursts from the beginning of the mission up until the present. Much of the data
were obtained with the Large Area Detectors, which have nearly all-sky coverage,
excellent sensitivity over two decades of energy and moderate energy resolution,
ideal for continuum spectra studies of a large sample of bursts at high time
resolution. At least 8 spectra from each burst were fitted with a spectral form
that consisted of a low-energy power law, a spectral break at middle energies
and a high-energy continuum. In most bursts (122), the high-energy continuum was
consistent with a power law. The evolution of the fitted high-energy power-law
index over the selected spectra for each burst is inconsistent with a constant
for 34\% of the total sample. The sample distribution of the average value for the
index from each burst is fairly
narrow, centered on $-2.12$. A linear trend in time is ruled out for only
20\% of the bursts, with hard-to-soft evolution dominating the sample (100
events). The distribution for the total change in the power-law index over the
duration of a burst peaks at the value $-0.37$, and is characterized by a median 
absolute deviation of 0.39, arguing that a single physical process is involved. We
present analyses of the correlation of the power-law index with time, burst
intensity and low-energy time evolution. In general, we confirm the general
hard-to-soft spectral evolution observed in the low-energy component of the
continuum, while presenting evidence that this evolution is different in nature
from that of the rest of the continuum.
\end{abstract}




%

\section{Introduction}

In the first six years of operation, the Burst and Transient Source Experiment
(BATSE), on board the {\it Compton Gamma Ray Observatory} (CGRO), has
accumulated a vast amount of spectral data on gamma-ray bursts. Although the
BATSE Large Area Detectors (LADs) have only moderate energy resolution compared
with the Spectroscopy Detectors (SDs), they have unprecedented effective area 
over their entire energy range (28 keV -- 1.8 MeV). By studying spectroscopy
data from the LADs for bright events such as those reported on by \cite{ford95},
who used SD data, we can track the evolution of fitted spectral parameters with
finer time resolution, and we can extend the analysis to fainter events. In this
paper, we analyze 126 bursts at high time resolution, with more than 8 spectra
per event, concentrating on the higher-energy behavior, where it was difficult
for the SDs to obtain good statistics.

As with much of the field of GRB studies, theoretical modeling of continuum
spectral emission naturally breaks into two periods: before and after the 
publication of the first BATSE results (\cite{meegan92}). The
paired observation of burst isotropy on the sky along with an 
inhomogeneous distribution of events with brightness, and presumably distance, 
has established the conclusion that GRBs occur much farther away, 
and are consequently much brighter, than previously expected. Instead of
comprising a nearby Galactic disk population, burst sources either reside in a 
very large Galactic halo or else they are truly cosmological (we will
not consider here another possible scenario: that bursts may arise in a local
heliospheric halo, such as the Oort cloud [e.g.: \cite{bickert94}; but also
see: \cite{cbt94}]). Such an uncertainty in distance has had dire theoretical consequences; no single model has surfaced
that can accommodate both distance scales, since such a model would have to 
account for luminousities that differ by $\sim 10$ orders of magnitude. The early 
theoretical work was dominated by the physics of strong-magnetic
field Galactic-disk neutron stars (see \cite{harding91}, for a review), which
has as its basis the efficient mechanism of quantum synchrotron emission. Of
course, energization of these systems was a crucial problem, in that the 
emission timescales are on the order of $10^{-17}$~s, for a typical field
strength of $10^{12}$~G required to produce a cyclotron absorption line 
fundamental at $\sim 20$ keV, as observed in X-ray pulsars (\cite{voges82}). 
Nevertheless, continuum modeling of then-current
spectral data enjoyed a moderate success (many references in \cite{ho92}). 

All this began to break down with the placement of burst sources no closer than
a large Galactic halo, as most of the strong-field models have restrictive
luminosity constraints. Cosmological burst emission scenarios proposed to date 
are less predictive, but have had little time yet to mature. For the most part,
interest has been focused on merging neutron stars, since the total energy
budget is about right for very distant events. What happens after the merger is
what distinguishes the models from each other. A simple fireball was 
proposed by many workers (\cite{cavallo78}, \cite{goodman86}, \cite{piran94}). 
Non-thermal emission, such as is observed in GRB spectra, is very difficult to produce in an optically-thick source, although as a fireball expands and becomes optically 
thin, a high-energy power-law component becomes possible. However, it was soon 
realized that in the environment of two
colliding compact objects, baryon contamination of the fireball would pose a
problem, diverting energy from the direct production of fireball radiation into the
acceleration of material (see discussion in \cite{fish95}). In order to address 
this problem, several workers proposed that the observed gamma-ray emission 
originates not in the original event but is a by-product of the kinetic energy 
gained at the expense of the fireball. Maximal acceleration of the explosion 
products leads to a relativistic blast front, which can cause shocks when colliding 
with interstellar material, either by encountering dense knots or eventually by 
sweeping up matter in the path of the shock front (\cite{rees92}, \cite{meszaros93}). 
Shocks can also arise internal to the outgoing relativistic wind, in the case where 
the central engine is variable (\cite{reesmesz94}, \cite{pacxu94}). It is important 
to note that the energy distribution of the 
shock-accelerated particles that gives rise to the observed emission is not 
predicted in any of these models; however, the distributions can be inferred from 
observation. The most efficient radiation mechanism is synchrotron, which 
produces a characteristic low-energy power-law behavior (\cite{katz94}, 
\cite{tavani96}). The high-energy spectral shape for this model
comes from the distribution of Lorentz factors for the baryons arising in the
shock, which is typically a broken power law. Dispersion of blast-front
velocities will give rise to observable hard-to-soft spectral evolution,
both in individual pulses, as well as over the course
of the entire event. Some of this behavior has been noted by \cite{ford95};
however, the opposite behavior is also seen, as well as a mixture of both.

Apart from the details of individual theoretical models, what new can be
learned from analysis of spectral data? First, we have the well-known 
observation that GRB spectra are non-thermal. There is good evidence that
some time-averaged GRB spectra are composed of power-law emission to
several 10s of MeV in energy (\cite{matz85}, \cite{hanlon95}). Burst
emission indeed reaches very high energies, as evidenced by the single
18 GeV photon observed by {\it EGRET} (\cite{hurley94}), albeit at a
considerable delay from the initial outburst. This alone can say much
about the distribution of particles doing the emitting, as well as the possible 
optical depth. Other than a multi-temperature blackbody, which can mimic
a power-law spectrum over a limited energy range, non-thermal emission
arises from non-thermal particles. The evolution of the particle
distribution, by cooling, for example, bears a simple relationship to
the evolution of the emission for many radiation mechanisms. In the
fireball model, optical depths are much greater than unity during the
phase in which the matter gets accelerated. Thermal emission from the
fireball is not observed in the gamma-ray band (although it may be
visible in X-rays below $\sim 20$ keV, \cite{preece96}). In any
cosmological model, it is very difficult to avoid conditions that will
rapidly lead to large optical depths via the photon-photon pair
production process. This occurs in the collision of two photons where
the product of their energies is greater than $2 m_{\rm e}^2/(1 - {\rm cos}\psi)$, 
$\psi$ being the angle between the photons' directions and
$m_{\rm e}=511$ keV is the rest mass of the electron. Many bursts have
substantial emission at 500 keV and greater, so if the high-energy
emission is not to be quenched by a runaway pair-fireball, the emission
must be highly beamed. The high energy power-law index and its time
evolution should constrain the mechanism through which particles are
giving up their energy in emission, as well as reflecting the behavior
of the injection mechanism. Cases where the high-energy component comes
and goes within a burst or is absent altogether may represent quenching
by a mechanism that rapidly increases the optical depth, such as
photon-photon pair production. In this case, it is expected that the intensity 
should drop during periods of quenched emission, or in other words, there should 
be a hardness-intensity correlation.

In this paper, we will present a study of the time-evolution of burst spectra, 
concentrating on the high-energy power-law component. In \S 2 we discuss the
burst sample selection and the details of the spectral fitting analysis. The
results are covered in \S 3 and their implications are discussed in \S 4. In 
the Appendices, we summarize the general characteristics of BATSE 
and then discuss in detail the energy calibration procedure for the LADs, 
which has made the current work possible.

\section{Analysis Methodology}

In order to have a sample of bursts with at least 8 spectra with count rates
high enough to obtain well-determined spectral parameters, we selected a subset
of bright bursts based upon either the total fluence or peak flux, as determined 
from the LAD 4-channel discriminator data (\cite{meegan96}). We required a fluence 
($> 20$ keV) greater than $4 \times 10^{-5}$ erg cm$^{-2}$. However, the set of 
bursts for which the fluence can be calculated is limited by
several considerations, such as data availability, telemetry gaps in the data
coverage and possible contamination of portions of some bursts with other active
sources (in particular, with solar flares in the first year of the mission).
Thus, we made an additional selection of those bursts which had a peak flux from
50 -- 300 keV on the 256 ms timescale in the 3B~Catalog (and later) above 10
photon s$^{-1}$ cm$^{-2}$. Each burst was then binned
in time, so that each spectrum to be analyzed had a signal to noise ratio (SNR)
of at least 45 in the typically 28 to 1800 keV energy range of the HERB data
(High Energy Resolution Burst data: for a description of the instrument and
spectroscopy datatypes, please see Appendix A). Bursts with less than 8 spectra
after binning were dropped from the sample. Most spectra in bright bursts
are well in excess of this SNR, which guarantees $> 2 \sigma$ of signal
per energy resolution element, assuming a flat count spectrum. Roughly 20 
resolution elements ($= \delta E$, the FWHM of the detector energy resolution) 
are required to cover the typical LAD energy range, thus the 128-channel HERB 
spectra are over-resolved in energy. LAD data types other than HERB are 
under-resolved, which is why HERB is preferred for spectroscopy. 
Some bursts did not have complete
coverage in the HERB data (especially before a flight software revision that
allowed longer accumulations during quiescent portions of a burst), in which
case we used other available data, as discussed below. There are 126 BATSE
bursts in our sample matching these criteria.

Background was determined independently for each channel, typically using
spectra from within $\pm 1000$ s of the burst trigger (giving at least three
background HER spectra before and after the burst). The form of the background
model was a fourth-order polynomial in each energy channel, where the fitted rates 
are time-averaged over each
spectral accumulation, rather than determined at the centers. This was done
to avoid underestimating the background rate at a peak or overestimating it at a
valley. The SNR was determined by comparison with the chosen background model, interpolated to the time of the accumulated spectrum. 

Spectra were fitted by one of several spectral forms, depending upon the best
fit obtained to the average spectrum over the entire burst. The primary spectral
form we used is the function of \cite{band92} (GRB, in table~\ref{table1}), 
which consists of two smoothly-joined power-laws: 
\begin{eqnarray}
f(E) & = & A (E/100)^{\alpha} \exp{(-E(2+\alpha)/E_{\rm peak})}\nonumber\\
{\rm if} \quad E & < & (\alpha-\beta)E_{\rm peak}/(2+\alpha){\rm ,}\\
{\rm and} \quad f(E) & = & A \{(\alpha-\beta)
E_{\rm peak}/[100(2+\alpha)]\}^{(\alpha-\beta)} \exp{(\beta-\alpha)}
 (E/100)^{\beta}\nonumber\\
{\rm if} \quad E & \geq & (\alpha-\beta)E_{\rm peak}/(2+\alpha){\rm .} \nonumber
\end{eqnarray}
The two power-law indices, $\alpha$ and $\beta$, are constrained such that
the resulting model is always concave downwards ($\alpha > \beta$; our definition 
includes a possible minus sign for each index). If, in addition, the
high-energy power-law index ($\beta$) is less than $-2$, the model peaks in $\nu
\cal{F}_{\nu}$ (that is, $E^2$ times the photon spectrum) within the BATSE
energy range. The model is parameterized by the energy of the peak in $\nu
\cal{F}_{\nu}$ ($E_{\rm{peak}}$), rather than the energy of the break between
the power laws ($E_0 = E_{\rm{peak}} (2+\alpha)/(\alpha-\beta)$). If the fitted 
value of $\beta$ is very negative, roughly less than 
$-5$, the spectral form approaches that of unsaturated inverse-Compton thermal
emission (\cite{randl79}), a low-energy power-law with an exponential cut-off 
(COMP, in table~\ref{table1}). This can be viewed as a generalization of the 
spectral form of optically-thin thermal bremsstrahlung (neglecting any Gaunt 
factor), which has a low-energy power law index of $-1$. The GRB spectral form 
is a continuous function that does not allow a sharp spectral break, so that 
in cases where $E_{\rm{peak}}$ ($< E_0$) is close to the high end of 
the energy range for the data, $\beta$ may not be well-determined. For such 
cases, we used instead a simple broken power law (BPL, in table~\ref{table1}), 
in order to force $E_{\rm{peak}} = E_0$, usually resulting in 
acceptable fits to the high-energy component. 

Since we are concerned in this paper about the high-energy power-law behavior,
we made a number of tests to be sure that our choices of spectral models do not
affect the end result. To do this, we fit several trial bursts with several
different models and compared the resulting fits. The simplest test of the
robustness of our fitting procedure was to fit a single power law to each
spectrum above a fixed cut-off energy that was determined by the maximum over 
the entire set of fitted spectra in the burst of the value of the break energy 
$E_0$ between the two power-law components. 
This eliminated any affect the fit to the low-energy data might have on the 
fitted value of $\beta$. That is, curvature in the global model fit 
may tend to pull the local fit of the high-energy power-law index to a larger 
or smaller value, depending on how well the actual data tolerate the curvature.
For example, the data may break more sharply than the model, which leads to
a fitted value for $\beta$ that is steeper than it should be. Conversely, in
the broken power-law model, with no curvature built in, the high energy power
law index may be pulled to a shallower value than the data require. Generally, 
the resulting time-history of the fitted parameters are consistent to 
within one-sigma errors. However, some differences were apparent when we
compared the time-histories of the fitted high-energy power-law component 
between these two models, when both were applied to the same burst, as can be 
seen in figure~\ref{beta_compare_fig}. The average values of the fitted power-law 
indices (weighted by the errors) over the entire burst were slightly different 
($\beta_{\rm ave} = -2.25$, for the GRB model fit; $= -2.16$, for the BPL), while
the underlying pattern of the time-history of the parameters were similar. So 
while the time evolution of the high energy portion of the spectrum could
be reliably traced by the fitted parameter for each model, there
remains some ambiguity in the average high-energy slope. This effect should be
worse for larger average values of $E_{\rm{peak}}$: the curvature inherent in
the GRB model tends to restrict the range of energies available for determining
$\beta$. The broken power-law model is plagued by a different problem: with an
energy resolution (FWHM) of approximately 20\% at 511 keV, we
usually cannot determine the exact position of the break energy using LAD count
spectral data. 

With the fitted values of the break energy and $\beta$ possibly closely correlated, 
the reported 1$\sigma$ error on each parameter is only part of the story. That 
is, the errors are most accurately determined from a multi-dimensional $\chi^2$ 
contour plot for the correlated parameters, as seen in figure~\ref{2dcontour}. 
The contours represent $\Delta \chi^2$ values appropriate for one parameter of 
interest, so that the 1$\sigma$ contour is at $\Delta \chi^2 = 1$ (for this 
figure only; usually, one would be interested in both parameters jointly, 
resulting in larger contours). The 1D 1$\sigma$ error limits are formed by the 
maximum and minimum of the error ellipse projected onto the axis of the parameter 
being considered. The actual 1$\sigma$ errors reported here are obtained from 
the diagonal elements of the covariance 
matrix for each fit; this is equivalent to $\Delta \chi^2 = 1$, with the 
additional assumption that the fitted parameter value lies in the center of 
the error interval. By taking into account the joint error between the 
parameters, $\Delta \chi^2$ is increased to 2.3, so that the fitted values of 
the high-energy power law index can be reconciled to within one or two sigma 
between the two different spectral forms.

The fact that we obtain acceptable fits with different spectral models reflects
on the ambiguity of the forward-folding process. Given the detector response, a 
count recorded in a given data bin could have come from a photon of any number of 
different energies, all greater than or approximately equal to the nominal energy 
range of the data bin. The dominant component of the response at low energies is 
the resolution-broadened photo-peak, centered on the photon energy. On top of this 
are counts derived from incomplete absorption of higher-energy photons in the 
detector, the off-diagonal component of the response. 
Consistent with the constraints imposed by the detector model, including
especially the energy resolution, a given photon model folded through the
detector response matrix will redistribute the predicted counts to best agree
with the observed data. Thus, the solution to the forward-folding spectral
fitting problem is not unique. 

Table~\ref{table1} summarizes global aspects of the fits performed for each burst. 
We use the 3B catalog name (\cite{meegan96}) and BATSE trigger number to
identify each burst, followed by the number of the detector with the smallest
zenith angle with respect to the source, the spectral model used for fitting, the 
number of fitted spectra, the time interval selected for fitting, the average 
of the fitted values for $E_{\rm{peak}}$ and the fluence, summed 
over the fitted spectra. In cases where there are two or more detectors reported 
in the third column, a summed 16 energy channel data type (MER) was used, usually
for lengthy events which ran out of HERB memory before the end of the burst. 
For a small number of cases where other data types were absent, we use SD 256 
energy channel data (SHERB); these are indicated in column three with an `S' 
appended before the detector number. The three models used in our analyses are 
indicated by their respective mnemonics (introduced above) in column four. 
The COMP spectral form has one less parameter than the others: there is no 
fitted high-energy power-law index. However, each of the models shares three
corresponding parameters: amplitude, low-energy power-law index and
$E_{\rm{peak}}$ (or spectral break energy for the broken power-law model). 
In the last two columns we indicate the average
value for $E_{\rm{peak}}$ in keV and the total fluence for the fitted interval
in erg cm$^{-2}$. Notice that three of the four bursts that required the COMP model
did so because the high energy power law was completely unconstrained; indeed,
for these bursts $E_{\rm{peak}}$ was also unconstrained, as the average value is
far greater than the energy of the highest channel available in the data 
(typically 1800 keV). In the following analyses, we shall exclude these four 
bursts, since no trend in the high-energy power law index can be determined 
with our data.

\section{Observations}

We should like to know several things concerning the behavior of the high-energy
power-law as a function of time. First of all, is it constant? If not, does the
index change smoothly with time, as with the hard-to-soft spectral evolution 
observed in the $E_{\rm{peak}}$ parameter
by \cite{ford95}? If the behavior is not smooth, is it correlated with other
observable features in the burst time history, such as the instantaneous flux or
the evolution of the low-energy spectral parameters? To investigate these questions,
we subjected the fitted values of the high-energy power-law index to several
statistical tests, and evaluated the probability the outcome of each could have
occurred randomly. The results of our analyses, shown in table~\ref{table2}, 
are described below. Each row of the table is indexed in the first column by the 
trigger name from table~\ref{table1}. For each burst, this is followed by the 
weighted average of $\beta$, the probability that a 
constant $\beta$ describes the data, the probability that a linear trend in 
$\beta$ describes the data, the slope from a linear fit to the time series of 
$\beta$, and the probabilities that the fitted values of $\beta$ are correlated 
with time, the burst time history or with the time series of $E_{\rm{peak}}$.

To start with, we would like to test the hypothesis that $\beta$ is a constant 
over the entire burst. In order to do this, we first computed a weighted average 
of the fitted values of the high-energy power-law index (which we will denote as 
$\beta$, regardless of which model we used for the fit) over the time interval 
selected for each burst. The weight applied to each term in the average is the 
square of the 1-sigma error, $\sigma_i$, of the fit:
\begin{equation}
\beta_{\rm{ave}} = \sum_{i} \biggl(\frac{\beta_i}{{\sigma_i}^2}\biggr) \bigg/
                   \sum_{i} \biggl(\frac{1}{{\sigma_i}^2}\biggr).
\end{equation}
In cases where the fit resulted in an undetermined value for $\beta$ for an
individual spectrum, the value was thrown out of the weighted average. It should 
be noted that, with weighting of the individual values, as well 
as the elimination of undetermined values, the result is different from the 
value of $\beta$ obtained from a fit to the integrated spectrum. The third column 
in table~\ref{table2} gives the probability for $\chi^2$ obtained by subtracting 
the weighted average from the actual fitted values in each burst. The 
$\chi^2$-values are calculated assuming the model, and thus a small value 
(such as $< 10^{-4}$) indicates a problem with the assumption and thus the 
likelihood that the model is false. A histogram of the logarithm of 
these probabilities in figure~\ref{ave_pl_prob} ({\it dotted line}), shows that 
for some bursts, at least, a constant 
$\beta$ is consistent. What is not shown are the 30 bursts for which the 
probability is essentially zero. Including the bursts for which the log. of the  
probability is less than $-4$, we have 42 out of 122, or 34\% of the total 
sample, that are not consistent with a simple, constant model in $\beta$. It is 
extremely unlikely that this distribution occurs randomly. 

The distribution of $\beta_{\rm{ave}}$, shown as a histogram in 
figure~\ref{beta_dist}, improves on earlier work by \cite{band93}, with a larger 
sample and better statistics per burst. However, the resulting values from these 
two studies cannot be compared directly, since here we have weighted each fitted 
value of $\beta$ by the parameter error, while in the previous study the fits were 
made to average spectra, which are implicitly weighted by intensity. Finally, 
the sample sets are different: the selection of events in \cite{band93} was based 
upon peak counts, not fluence or peak flux, since these were unknown at the time. 
The median value for the sample is $-2.12$, with an absolute deviation width of 
$w_{\rm ADev} \equiv {1 \over N}\,{\sum_{j\,=1}^{N} \mid x_{j} - x \mid}$ 
= 0.23 (where $x_{\it med}$ represents the median, which minimizes the absolute 
deviation), compared with the standard deviation of 0.30. The distribution has an extended negative tail that gives it a skew value of $-0.73$ (the skew is defined 
as the dimensionless third moment of the distribution, and is 0 for a Gaussian), 
large compared with the expected standard 
deviation of the skew of $\sqrt{15/N}=0.35$ for a purely Gaussian distribution. 
Given the large variation of other spectral parameters, such as $E_{\rm{peak}}$ 
which has a distribution at least as wide as the range of possible values, it is 
surprising that the high-energy behavior is so 
restricted. Plotted over the total distribution in figure~\ref{beta_dist} is 
a histogram of those bursts for which $\beta$ is consistent with being constant 
(log. probability $> -4$ from figure~\ref{ave_pl_prob}).

Obviously, a constant value of $\beta$ is not acceptable for many bursts. A clear 
example of this is presented in figure~\ref{beta_1085}, which shows the time 
history of $\beta$ during 3B911118 and is an example of general hard-to-soft 
spectral evolution in $\beta$. The Spearman rank-order 
correlation of $\beta$ with time is given in column 6 of table~\ref{table2}. 
The correlation coefficient {\it r} is distributed between $-1$ and 1, and can be 
converted through the combination
\begin{equation}
t = r \sqrt{\frac{N - 2}{1 - r^2}}
\end{equation}
to a Student's {\it t}-distribution for $N - 2$ degrees of freedom. Unlike the 
$\chi^2$ probabilities, correlation coefficients that are not consistent with 
roughly a normal distribution around 0 reject the null hypothesis that no correlation 
exists; therefore, small probabilities indicate significant correlation. The probabilities associated with {\it r}, calculated using equation 3 along with the 
number of spectra fitted ($N$) from column 5 of table~\ref{table1}, reveal that 
a trend in the data exists for at least 21 of the events at the 
$10^{-3}$ significance level or smaller. This is a robust estimator for 
correlation; it indicates when a correlation is almost certainly present. 
However, the Spearman test does not take into account the errors for each point, 
so if there are a large number of outliers with large errors in the sample, the 
test will come up with poor results. Figure~\ref{corr_dist} presents the distribution 
of the time correlation coefficients ({\it solid line}). The bulk of the distribution 
consists of negative correlations, indicating an anti-correlation of the power-law 
index with time, or hard-to-soft spectral evolution.

A linear fit to the time history of $\beta$ also indicates whether there is a 
monotonic trend 
in the data, while accurately treating the errors in the fitted power-law indices. 
The fifth column of table~\ref{table2} gives the linear coefficient, or slope, of 
such a fit, having the units of change in $\beta$ per unit time, or s$^{-1}$. 
The sign is such that hard-to-soft spectral evolution ($\beta$ grows more 
negative in time) results in a negative slope. The $\chi^2$ probability for this 
fit is given in the fourth column of the table and the distribution is also plotted 
on figure~\ref{ave_pl_prob} ({\it solid line}). In 24 cases out of the total sample, 
the log.\, probability was less than $-4$, indicating that the linear trend was a poor 
model of the data for those events. Comparing this result to that for the model of 
constant $\beta$, however, more bursts had acceptable fits to a linear trend 
at the same significance level (98 compared with 80 out of 122). There are far more 
cases of hard-to-soft spectral evolution (100) than there are for soft-to-hard evolution, which was already evident in figure~\ref{corr_dist}. The 
first spectrum in many bursts is the hardest (see figure~\ref{beta_1085}), 
while at the same time being one of the weakest. Since each burst has a different
duration, the slopes in physical units may not be directly comparable. However, the 
fitted slope in $\beta$ times the duration of the fitted time interval, from column 
6 of table~\ref{table1}, is a dimensionless parameter ($\Delta \beta$) that 
represents the total change in $\beta$, assuming that the evolution in $\beta$ is 
linear (as it is for the majority of the sample). Figure~\ref{slope_dur} shows that 
the distribution of $\Delta \beta$ has a single, roughly symmetric peak centered 
on $-0.374$, with one outlier (not shown in the figure). The median absolute
deviation width of the distribution is $w_{\rm ADev} = 0.392$, compared with a 
standard deviation of $\sigma = 0.516$. This argues that a single physical process  characterizes the majority of the 
sample; and again points out that hard-to-soft spectral evolution is typical 
behavior for the high-energy power-law component. Physical mechanisms for burst 
energetics should account for this, possibly via depletion of a reservoir of 
energy that is available for the burst. Otherwise, it may be that when the 
high-energy portion of the emission changes beyond this point, the total 
emission is quenched.

The linear fit to the power-law indices does not characterize the distribution 
well for many bursts (24 out of 122), indicating that other types of behavior may 
be present. 
Figure~\ref{beta_1085} serves as an example of a burst that has strong hard-to-soft 
spectral evolution but where the linear fit is unacceptably poor. The residuals to
the fit have considerable scatter that is correlated in successive time bins 
in several places on the figure. It is these residual patterns that we are 
interested in. Two possibilities are easily tested: there may be a correlation
between the high-energy behavior and intensity within a burst (clearly not the
case for 3B~911118 in figure~\ref{beta_1085}), or the high-energy spectrum may be correlated with the evolution of the low-energy spectrum. Burst 3B~911118 is an 
example of this behavior, as can be seen in figure~\ref{epeak_beta}, where the 
fitted values of $E_{\rm peak}$ (representing the low-energy behavior) and 
$\beta$ have been plotted against each other.

For the case of correlation between hardness (as measured by the high-energy
power-law index) and intensity (measured as total count rate in the fitted
energy interval: $\sim 28$ -- 1800 keV), we applied two statistical tests to the
data and multiplied their probabilities in order to screen for candidates. The
tests (described below) are likely to be correlated; however, each measures the
hardness-intensity correlation differently, so that their product combines the
best of each. We set the threshold for significance at $10^{-6}$ for the
product, so to avoid false positives as much as possible. In both tests, we
removed the first-order trend in the data by dividing by the linear fit to the
power-law indices (which is described above). We do this, despite the fact that
many bursts don't show a linear trend in the high-energy power-law index, since
there are a considerable number of bursts that do have a significant correlation
between $\beta$ and time, while the burst intensity manifestly does not: a typical 
burst will have overlapping regions of both positive (rising portions) and negative 
(falling) correlation with time, so that the whole ensemble of $\beta$ values has 
no correlation. The 
overall linear trend may be larger than the amplitude of the residuals of the 
fitted linear model (this is the case in figure~\ref{beta_1085}), in which case  
there is no significant hardness -- intensity correlation as determined by $\beta$ 
alone. After detrending, the residuals may or may not be correlated with
intensity. The Spearman rank-order test is relatively unequivocal: that is, if
the resulting probability is low enough, then the desired correlation definitely
exists. However, the converse is not true: the test can fail badly since it
ignores the one-sigma errors in the fitted power-law indices. For this reason,
we also have calculated the linear correlation coefficient between the detrended
values of $\beta$ and intensity, where the inverses of the variances on the
detrended power-law indices are used to weight their contribution
(\cite{press92}). For this case, individual, poorly-determined indices that are
only a few sigma away from being consistent with correlation contribute the same
as well-determined ones closer to the center of the distribution. In practice,
while this kind of test is a poor indicator of whether an observed correlation
is statistically significant, it is a rough indication of the strength of a
correlation under the assumption that a correlation definitely exists, so the
two statistical tests we've chosen complement each other, to a certain extent.
Their product selects those bursts that have low probabilities (indicating
strong correlation) from both tests (assuming that by detrending no significant
correlation was introduced that was not present in the original data). We have
indicated the combined probabilities from both tests in the seventh column of
table~\ref{table2} and also indicate the sign of the linear correlation
coefficient. Since the power-law indices were detrended, a positive sign
indicates a negative actual correlation; that is, the high-energy behavior is
opposite that of the burst time history. An example of positive correlation in
the detrended values of $\beta$ for 4B~960924 is shown in
figure~\ref{detrended_beta_5614}. A small number of bursts (9), have
significances less than $10^{-6}$. Of these, 6 are examples of positive
correlation. A larger number (24) are significant at the $10^{-4}$ level.

Another possible type of behavior in $\beta$ that is testable with our data is a
correlation with the low-energy spectral evolution. The most obvious such
behavior is the hard-to-soft spectral evolution of $E_{\rm{peak}}$, discussed by
\cite{ford95}. $E_{\rm{peak}}$ is a good measure for overall spectral evolution
since it marks the peak in the power output of the spectrum per log. decade. Of 
course, $E_{\rm{peak}}$ is not defined for those portions of a burst where 
$\beta > -2$; in that case, we substitute the break energy of the spectrum instead. 
In addition, we wish to check for higher moments of correlation than is possible
with a linear trend of $\beta$ in time, which was discussed above, such as the
evolution of $\beta$ within individual peaks of a burst. In table~\ref{table2},
column 8, we calculate the Spearman rank-order probability that the distribution
of $\beta$ for a given burst is correlated with the distribution of
$E_{\rm{peak}}$, which stands in here for the low-energy behavior. The best
example of correlation with $E_{\rm{peak}}$ is shown in figure~\ref{epeak_beta},
which is a plot of the two fitted parameters against one another for 3B~911118.
Out of 122 bursts, 15 bursts have probabilities less than $10^{-3}$, indicating
correlation, and out of these only 5 have significant hard-to-soft spectral
evolution, as measured by how many sigma the slope in $\beta$ in
table~\ref{table2}, column 5, deviates from 0. The important point is that,
whereas hard-to-soft behavior can be demonstrated for large numbers of bursts in
the evolution both $E_{\rm{peak}}$ and $\beta$, this behavior is generally not 
correlated between the two.
Indeed, hard-to-soft evolution of $E_{\rm{peak}}$ within individual peaks of
a burst is not typically observed with $\beta$, otherwise, far more instances of
correlation between the two would have been observed.

\section{Discussion}

In this series of BATSE spectral analysis papers, we have demonstrated several
times the universal suitability of the `GRB' spectral form for fitting burst 
spectra, whether it is applied to the total spectrum averaged over the burst 
(\cite{band93}), to time-resolved spectroscopy of bright bursts in the SD data 
in \cite{ford95}, to joint fits of time-averaged spectra of bright bursts with 
the low-energy discriminator data (\cite{preece96}; although we see the 
model break down with low-energy excesses observed in 15\% of GRBs) and now to 
time-resolved spectroscopy of bursts observed mostly with the BATSE LADs. In 
figure~\ref{beta_dist}, we now see that there is evidence of an average high-energy 
power law index that is $\sim -2$ in a large number of GRBs. In 
addition, the variance of this index over the sample is similar to that obtained 
by \cite{pend94a}, using BATSE LAD discriminator data. 

Table~\ref{table2} presents evidence that $\beta$ is not constant for 42 out 
of 122 bursts in our sample. The typical 
change in $\beta$ over an entire burst, $\sim 0.4$ (figure~\ref{slope_dur}), is 
small compared with the average value of $\beta \approx -2.1$. We should 
consider which of the many emission models proposed for GRBs are consistent with 
these observations. A $-2$ power law slope is evidence for single-particle 
cooling, from either synchrotron losses or Compton scattering 
(\cite{blumgould70}). Typically, one would integrate the energy loss
rate over the particle distribution; however, particles that are relatively cool
with respect to their large, possibly relativistic bulk motion can be treated as
monoenergetic in interactions with static external particles or fields.
Bremsstrahlung losses are another matter. Such scenarios have been proposed for
bursts of cosmological origin for external shocks, (\cite{rees92} \& 
\cite{meszaros93}) as well as for synchrotron shocks (\cite{katz94} \& 
\cite{tavani96}). It should be noted that the cooling timescale for most
expected processes, especially those like synchrotron that involve magnetic
fields, is far shorter than observed burst lifetimes by many orders of
magnitude. In fact, this is a common problem with GRB models: an
unspecified energy storage mechanisms usually must be invoked in order to extend
the emission. Relativistic bulk motion, which is necessary to ensure
that bursts do not degenerate into a pair fireball, can multiply the lab-frame
lifetime by the Lorentz factor, usually considered to be on the order of 1000.
This is not nearly long enough for processes such as synchrotron emission whose 
characteristic timescale may be on the order of $10^{-17}$~s. Clearly, in bursts, 
there is a reservoir of energy, possibly the protons that carry the bulk of the 
kinetic energy in the blast wave.

It appears that hard-to-soft spectral evolution predominates over soft-to-hard,
as observed already in \cite{norris86}, \cite{ford95} and \cite{band97}. 
In our study, the high-energy behavior
follows this trend at the greater than the $3 \sigma$ level in 50 out of 122
cases, while the opposite is true for only 5 bursts at the same significance.
This is independent of the low-energy behavior; indeed, we have a significant
correlation with the low-energy behavior in only 15 cases and out of these, 5
have significant hard-to-soft spectral evolution. Taken together, we have
evidence that the high-energy behavior is very much independent of the rest of
the spectral evolution of a burst; in 35\% of the cases, there is hard-to-soft
spectral evolution, and no evolution in most of the rest, only 10\% of all
bursts failing the linear fit $\chi^2$ test.

As seen in figure~\ref{beta_dist}, there is a small group of `super-soft' bursts
characterized by $\beta_{\rm ave} < \sim -3.0$. Along with 4B~970111, which was an 
extremely bright burst with no apparent high-energy power law component (it was fitted 
with the COMP model), we have three such events. Several of these have no
detectable emission above $\sim 600$ keV. This behavior is similar to the `no
high-energy' bursts of \cite{pendleton97}, which were shown to be homogeneous in
space. Since most of the homogeneous bursts were relatively weak, compared with
the entire sample, here we must be observing the brightest few of that set,
rather than 20\%, as reported in \cite{pendleton97}. There may actually be a
continuum of burst properties, with these bursts representing the furthest
extreme. Bursts in this extreme (as well as some portions of other bursts that
have very steep high-energy power laws) may be an indicator that some
emission-limiting phenomenon such as pair-plasma attenuation may be at work.
Indeed, in many cases, spectra in these bursts can be fitted by a spectral form
that does not require a high-energy power-law (such as the COMP model). This
also fits in with the observation that such events are typically weaker than
average. In the context of shock models of GRBs, several parameters of the
particle energy distribution determine the resulting spectrum. These may be
factors such as the shape of the distribution, whether it is a power law, the
maximum energy or the bulk Lorentz factor. It is very likely that the maximum
energy of the accelerated particle distribution resulting from the shock could
be drawn from an enormous range (out to several GeV, at least), depending on the
conditions at the shock. Thus, the super-soft bursts may be representative of
particle distributions that arise from weak shocks, affecting the shape or
maximum energy in such a way to limit the high-energy emission.

\section{Summary}

In this study, we have looked in detail at the temporal behavior of the
high-energy power-law portion of GRB spectra from a sample of 126 bursts
selected by either high flux or fluence. The average over all fitted spectra of
all bursts in the sample for the high energy power law index ($\beta$) is $\sim
-2.12$, although fitting a constant, average index to the time history of
$\beta$ in each burst resulted in unacceptable $\chi^2$ values for 34\% of the
bursts. In addition, of those bursts in which $\beta$ is not constant, a large
number (100) show hard-to-soft spectral evolution, compared with those that have
an overall, significant soft-to-hard trend. The total change of $\beta$ over the
time interval chosen for fitting has a single-peaked distribution, centered on $-0.37$,
indicating that theoretical modeling will have to explain why most bursts favor
this value. In several bursts, the hard-to-soft spectral evolution is correlated
with similar behavior at lower energies. We also find that some bursts have a
significant correlation between $\beta$ and the burst time history, or
equivalently, instantaneous flux. Some bursts in the sample were too soft to be
characterized by a high-energy tail, while there are intervals in many bursts
that have similar behavior, as has been reported by \cite{pendleton97}. Taken 
together, these results show that the high-energy spectral component has a rich 
life, independent to a large extent of the behavior of much the rest of the spectra.

\acknowledgments

Many thanks to Surasak Phengchamnan and Peter Woods for generating a 
list of post-3B catalog fluences and peak fluxes. We also thank the anonymous referee, 
for comments that lead to improvements in the paper. This work could not have been 
possible without 
our spectral analysis software (WINGSPAN). It is publicly available from 
the BATSE webserver: http://www.batse.msfc.nasa.gov/. BATSE work at UCSD is supported
under NASA contract NAS 8-36081.

\appendix
\section{BATSE Large Area Detectors}

The BATSE LADs are a set of eight identical NaI detectors, which are mounted on
the corners of the {\it CGRO} and oriented to ensure maximum all-sky exposure.
Perhaps the most important feature of the BATSE instrument is its ability to
localize a transient cosmic source by the comparison of counting rates in the
four detectors that directly see it (\cite{pendleton96}). This is an invaluable
aid to spectroscopy, since the detector response is a strong function of the
source-to-detector axis angle, with differing responses at different energies 
(\cite{pend95}). Thus, without location information, the detector response cannot 
be fully modeled, and spectral model fitting cannot be done accurately. 

Spectral data from the LADs are compressed to either 128-channel high
energy resolution background and burst data (HER and HERB datatypes,
respectively) or 16-channel continuous background or medium energy resolution
burst data (CONT and MER datatypes). The HER background data are typically
accumulated over 300 s, while the CONT data are always accumulated every 2.048
s. The HERB burst trigger data are accumulated in a time-to-spill mode: one 
spectrum is generated in the time it takes to record 64 k counts (in units of 64 
ms), currently with a fraction of the last available background rate subtracted, 
to ensure that longer accumulations are taken over periods when the burst has
returned to background levels. This fraction was zero for roughly the first half
of the mission, so bright, highly-variable bursts commonly ran out of available 
memory. For the four detectors recording the highest count rates at the time of 
the trigger, there are 128 spectral accumulations, each 128 ms in duration or 
greater. The lowest seven channels of the 128 are at or below the analog
lower-level discriminator (LLD), and are unusable; the highest few channels suffer
from saturation in the pulse amplifier and thus are also thrown out. The
remaining channels are spaced quasi-logarithmically in energy, falling between
approximately 28 keV and 2 MeV, with the exact energy coverage of each channel
in each detector determined by a channel-to-energy conversion algorithm. It is 
important to note that these energy ranges are quite stable through the mission, 
due to automatic gain control of the PMT voltages. The energy resolution of the 
LADs was measured on the ground to be $\sim 20$\% at 511 keV (\cite{horack91}), 
and has been quite stable in orbit.

\section{Energy Calibration Methodology}

In order for spectroscopy to be possible with the LAD HERB data, we have had to
apply a correction to the channel-to-energy conversion algorithm that was
developed before the launch of the spacecraft. Measurements of several
calibration sources at known energies resulted in an empirical relationship
between channel number and channel energy threshold (\cite{lestrade91}). 
The function fitted was essentially linear, with a small
non-linear term (significant only at low energies), proportional to the square
root of the channel number; thus there are three fitted parameters. After
several bright bursts were observed in orbit, it became clear that each detector
had a systematic pattern of residuals, localized to the low end of the count 
spectrum. With the assumption (tested below) that these features are intrinsic to 
each detector, and not a function of detector-to-source angle or source intensity, 
we developed a method of calibration using in-orbit data.

In order to properly calibrate the detectors, we must choose bright objects with
well-known spectral properties, seen by each detector. Solar flares are generally
not usable, since they are rarely seen by half the detectors, due to the pointing
constraints of the spacecraftÕs solar panels, and their spectra are typically too
soft. Earth occultation data from the Crab nebula was used by \cite{pend94b} to
calibrate the 16 channel CONT spectra. However, this was not feasible for HER
spectra because of telemetry constraints. We are left with bursts themselves.
Averaged over their entire time history, at least some bursts can be expected to
have a fairly smooth spectrum (\cite{band93}). Spectral features, such as lines,
will tend to average out over time and in the LAD data will not contribute much
overall, due to the moderate energy resolution of the detectors. 
For bright bursts, we can
precisely determine the average spectrum from the well-calibrated SD spectral
data (\cite{band92}) to use as a constraint in a joint fit with the LAD spectral
data. The single time-averaged spectrum from the calibration burst is no longer
available for spectroscopy; however, individual spectra from the burst are still
usable for our analyses, for two reasons. First, continuum spectral fits are
robust, in that they sample broad features in the spectrum, rather than the
behavior of individual channels. Second, the calibration affects only the lowest
channels of the spectra, and therefore does not affect spectral fitting of the
high-energy power-law index, as long as it is determined by counts above $\sim$
150 keV. In the present paper, we needed to obtain a global fit to each spectrum,
so it was important to calibrate the lower channels as well as possible.

The general process is iterative: we jointly fit the LAD and SD spectral data
for an entire outburst interval in a bright burst, using the standard
calculation for the LAD data energy thresholds. The residuals of the fit to the
LAD data are used to determine by how much to adjust the energy of each data
channel edge in order to bring the count rate closer to the model rate. With this
new set of edges, a new detector response matrix (DRM) is generated to account
for the shift in the position of the photopeak with the change in output edges
and the accompanying change in total response. The photon model is recalculated
with the new DRM and count rate residuals are again determined. Since the
pre-flight calibration produces acceptable agreement above $\sim 150$ keV, we 
limited the re-calibration to energies below 150 keV. We also enforced a fixed 
lower energy for HERB data channel 7 ($\sim 25$ keV), to limit the corrections to 
apply only to channels above
the energy of the LLD, as this is currently not modeled in the DRM. The freedom of
lower-energy edges to wander is highly constrained in the joint fit with the SD
data, which overlap the LAD energy range and can extend the continuum fit to 
lower energies by up to 10 keV. Each of the edges within the two limits 
are recalculated in each new cycle until the value of $\chi^2$ for the fit
stops decreasing. For each of the eight detectors, one calibration burst yields a 
set of offsets of new edges relative to the original edges, which can then 
be applied to all bursts
observed by that detector throughout the mission. We have extensively tested the
hypothesis that the non-linearities are intrinsic to the detector by examining
the residuals to spectral fits of several very bright bursts in each detector
with the new calibration. We have found excellent agreement of the calibration
results between bursts, regardless of the angle, intensity or hardness of any
given event.

\clearpage

\figcaption[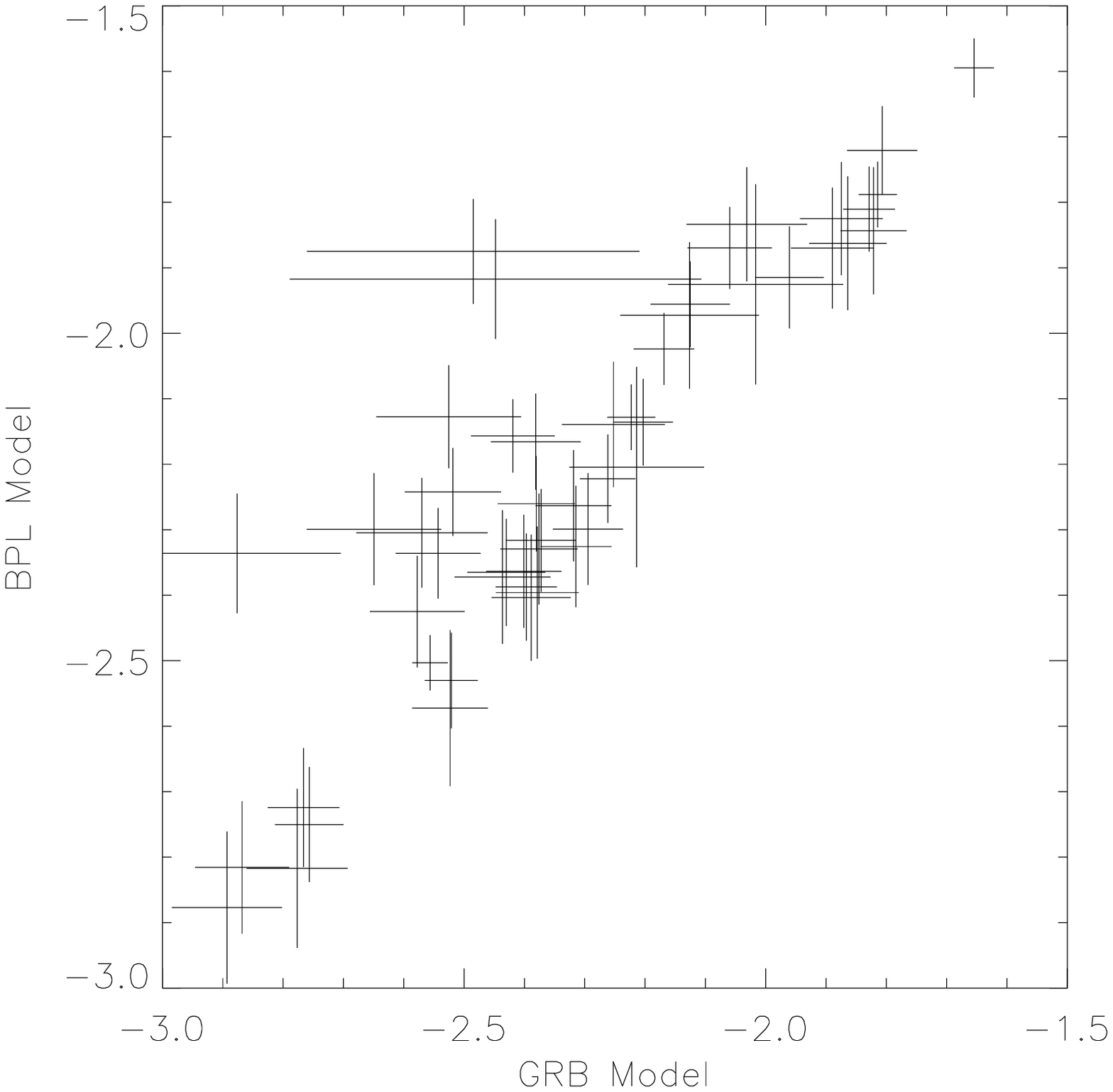]{A comparison between the fitted values for 
$\beta$ from two different spectral models (GRB \& BPL) for 4B 950403. 
\label{beta_compare_fig}}

\figcaption[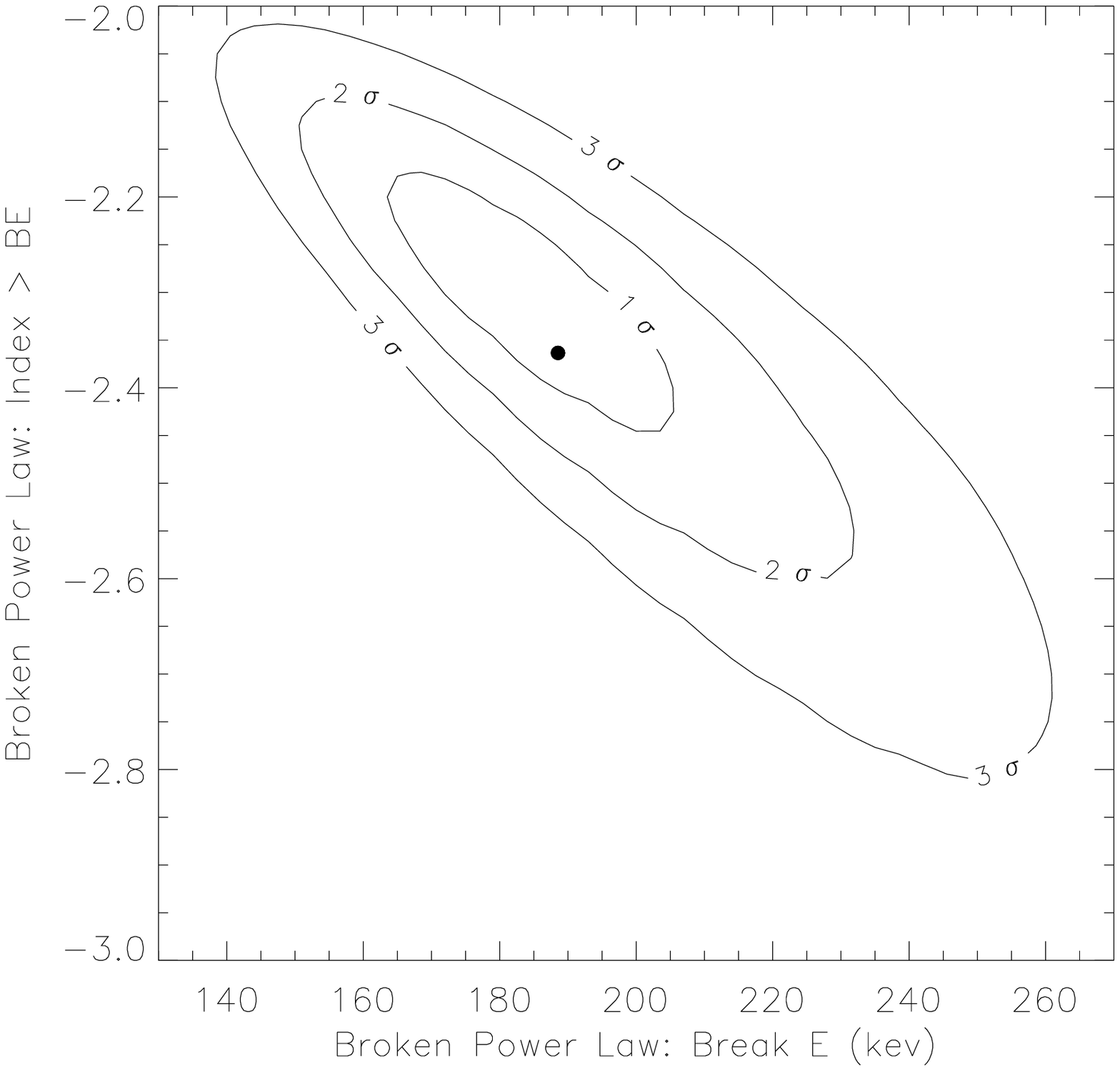]{A 2-D contour plot showing correlation between 
$E_{\rm peak}$ and $\beta$ in the broken power-law model. A solid dot indicates 
the best-fit values for the two parameters. As discussed in the text, note that 
the $1\sigma$ contour on this 
figure is appropriate for only one parameter of interest, that is, it represents 
$\Delta \chi^2 = 1$, rather than 2.3, which would contain 68\% of the {\it joint} 
probability. The data are from the interval 7.680 -- 7.808 s of 4B 950403. \label{2dcontour}}

\figcaption[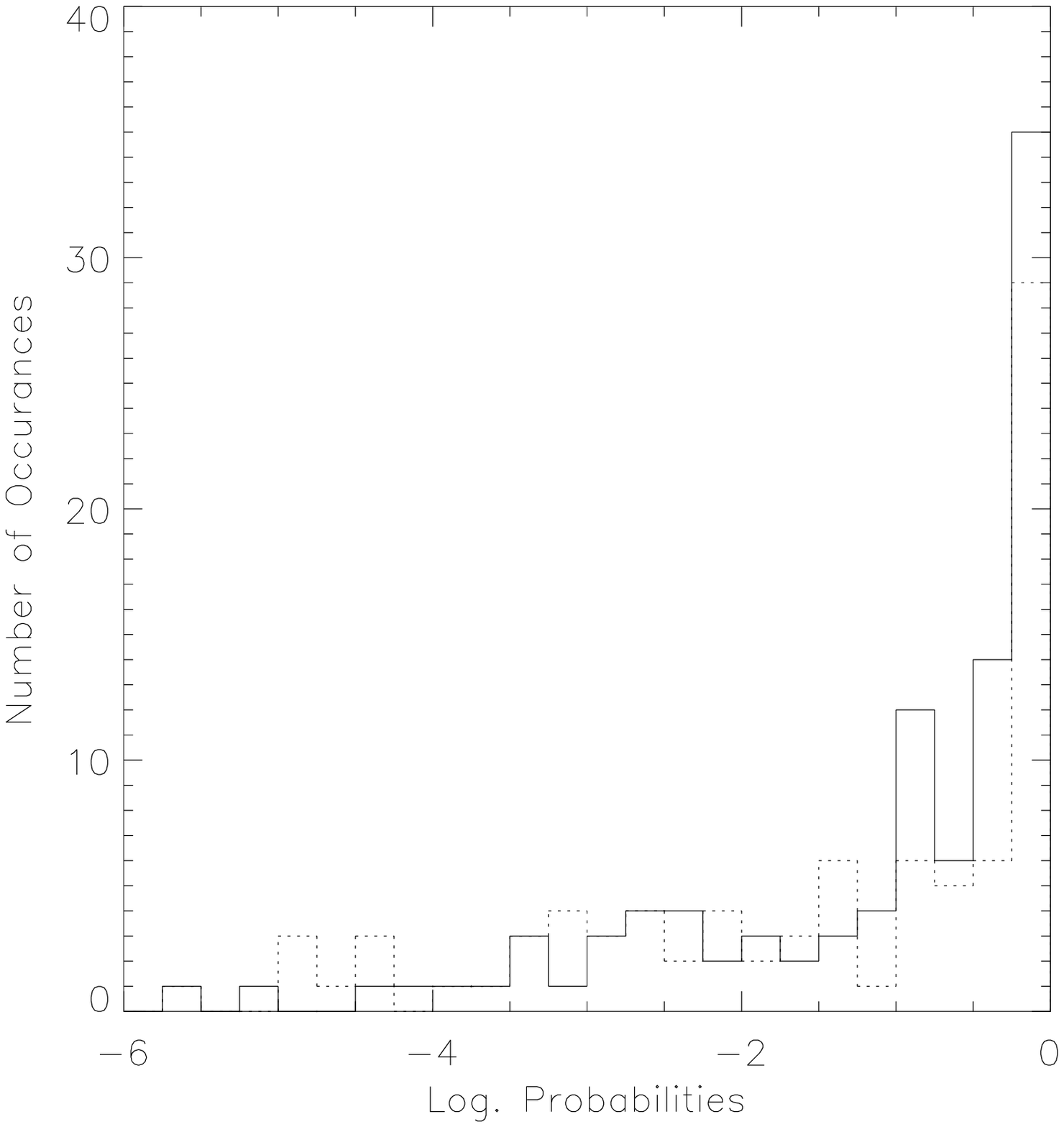]{Histograms of the log. of the $\chi^2$ probability 
that $\beta$ is a constant ({\it dotted line}) or exhibits a linear trend ({\it 
solid line}) throughout each burst. The bursts that have probabilities consistent 
with zero (indicating rejection of the model) are not shown for either of the two 
distributions (34 and 20, respectively). \label{ave_pl_prob}}

\figcaption[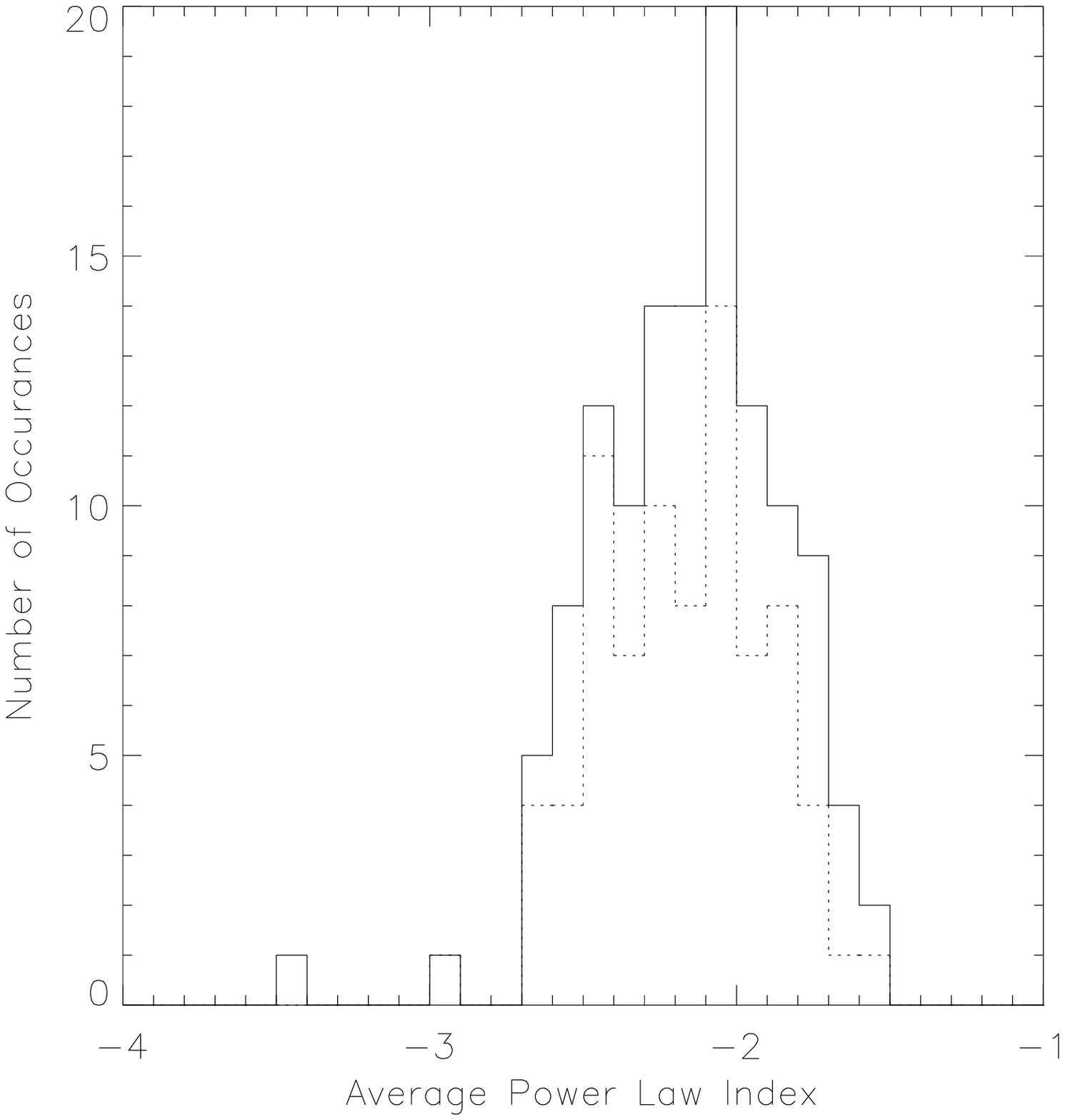]{Histogram of the weighted average of $\beta$ for
the burst sample ({\it solid line}). Overplotted is the subset of bursts for which 
a constant value for $\beta$ resulted in acceptable values for $\chi^2$, at the 
$10^{-4}$ level of significance ({\it dotted line}). \label{beta_dist}}

\figcaption[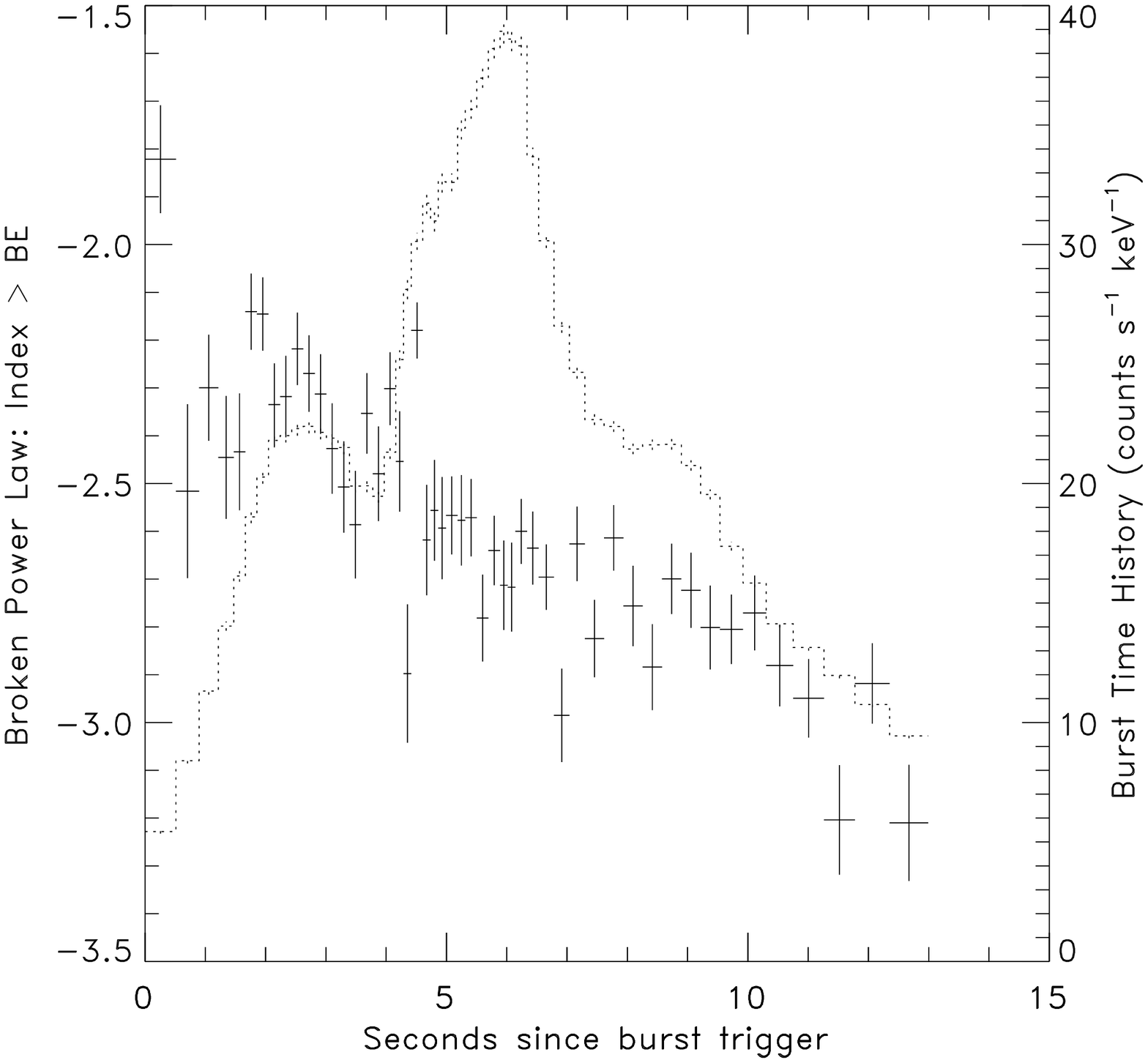]{Example of hard-to-soft spectral evolution in 
$\beta$ for 3B~911118. 
Both the fitted value of $\beta$ with their errors ({\it solid crosses}) 
and the burst count rate history ({\it dotted lines}) are plotted as a 
function of time. \label{beta_1085}}

\figcaption[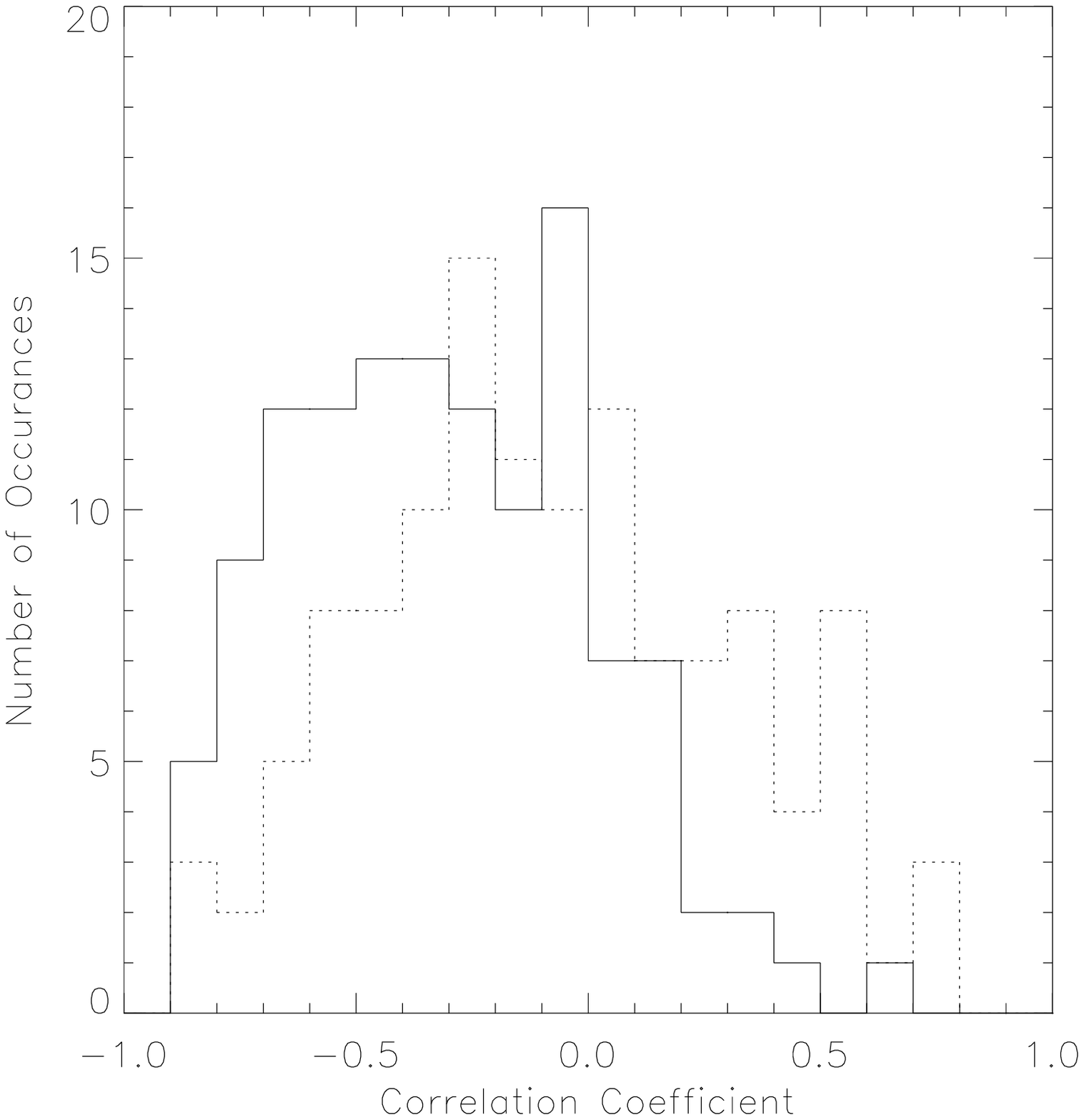]{Histograms for the distributions of coefficients of correlation between $\beta$ and time ({\it solid line}) and $E_{\rm peak}$ ({\it 
dotted line}) for the burst sample. \label{corr_dist}}

\figcaption[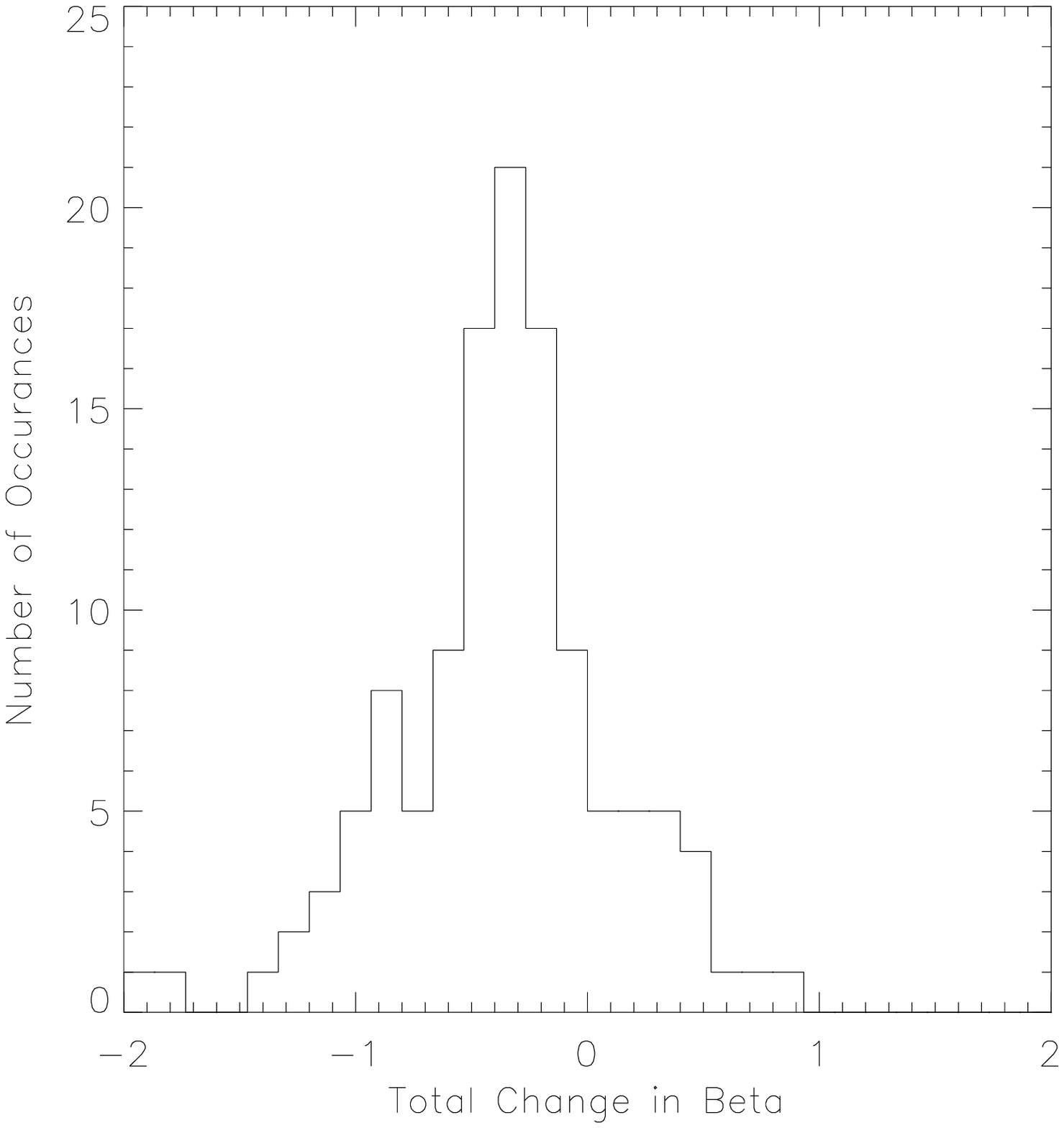]{Histogram of the total change in the high-energy 
power-law index (fitted value of $d\beta / d{\rm t}$ times the total 
time interval) for each burst in the sample. To improve 
the clarity of the figure, one outlier is not shown. \label{slope_dur}}

\figcaption[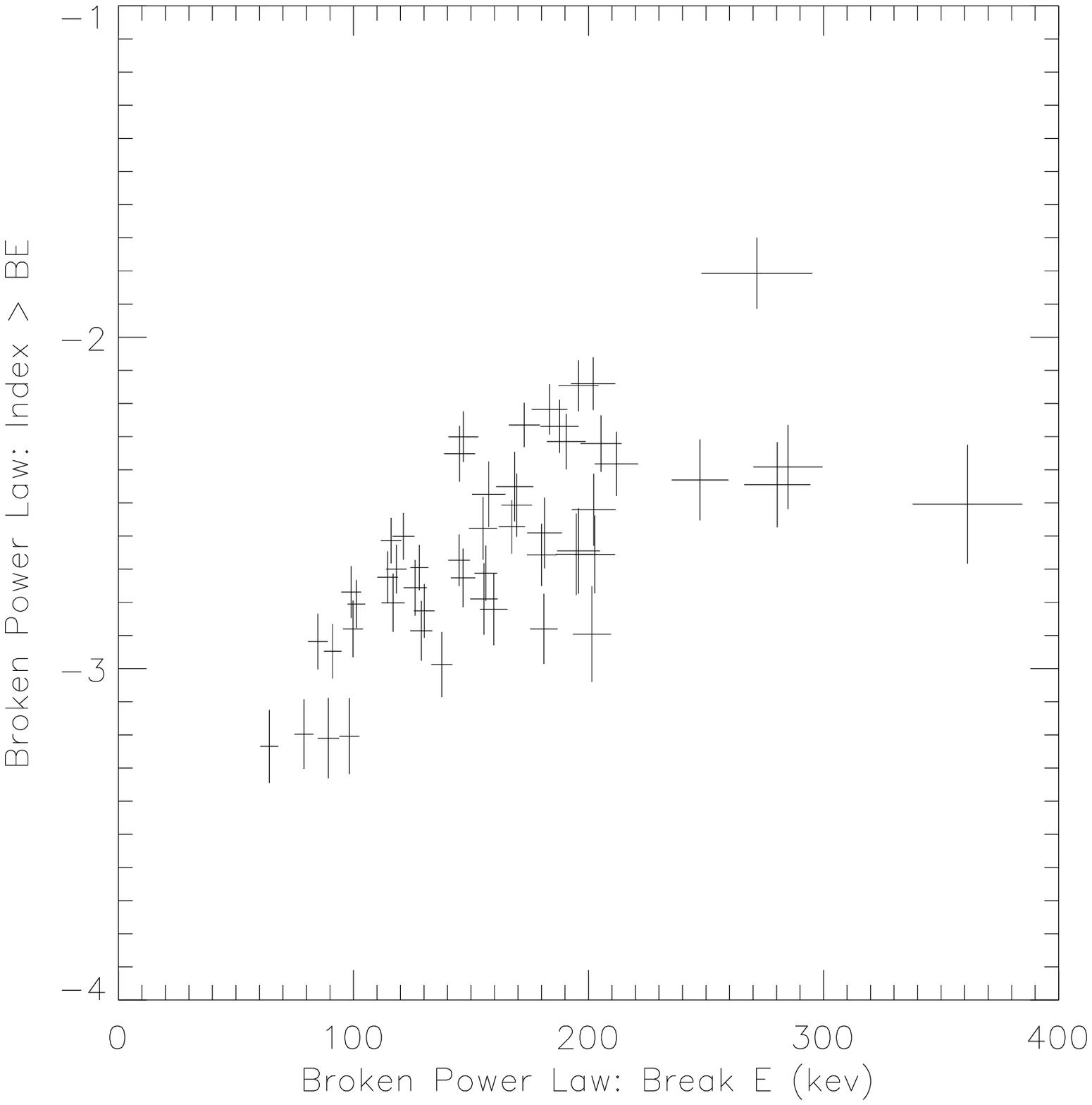]{The fitted values of $E_{\rm peak}$ plotted against 
$\beta$ for 3B~911118, illustrating correlation between the time evolution
of both hard and soft spectral components in a burst. \label{epeak_beta}}

\figcaption[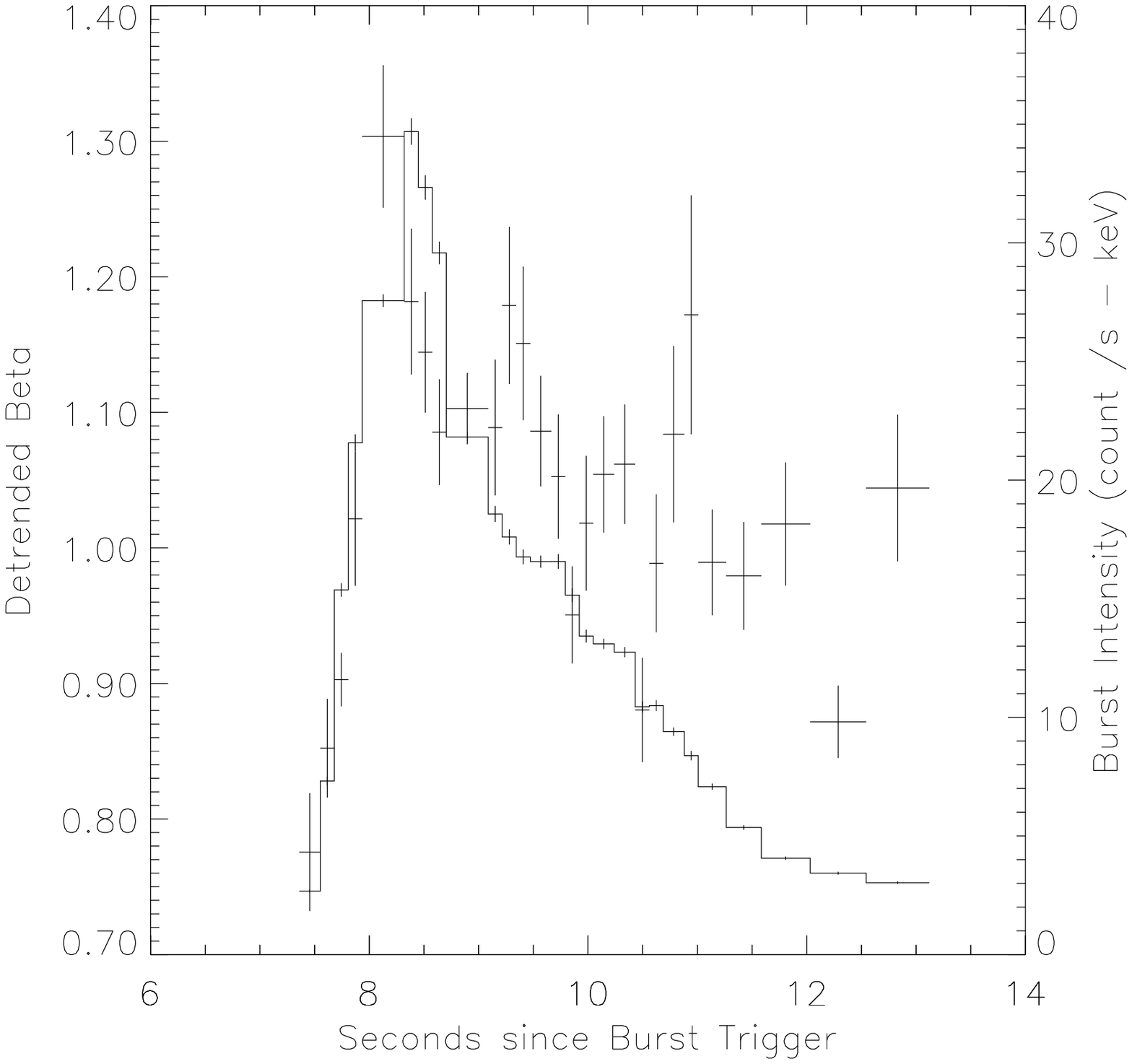]{Example of positive correlation between fitted  values of $\beta$ divided by their linear trend (`detrended' $\beta$ -- 
{\it crosses}) and the time history of the count rate for 4B~9960924 (the count rate 
has been divided by the energy range $28 - 1800$ keV -- {\it 
solid histogram}). \label{detrended_beta_5614}}

\dummytable\label{table1}
\dummytable\label{table2}



\clearpage

\end{document}


\begin{deluxetable}{crcrcrrr}
\tablenum{1}
\tablewidth{0pt}
\tablecaption{General Characteristics of Sample Bursts. \label{table1}}
\tablehead{
\colhead{Burst}   & \colhead{Trigger}   & \colhead{Detector} & \colhead{Model}  & 
\colhead{No.} & \colhead{Time} &
\colhead{Average}     & 
\colhead{Fluence}  \nl
\colhead{Name\tablenotemark{a}}   & \colhead{No.}   & 
\colhead{No.\tablenotemark{b}} & 
\colhead{Used\tablenotemark{c}}  & 
\colhead{Spectra} & \colhead{Interval ($s$)} &
\colhead{$E_{\rm{peak}}$ ($keV$)}     & 
\colhead{($erg~cm^{-2}$)\tablenotemark{d}}  
}
\startdata
3B 910421&   105&  7&   GRB&  14&  6.4e$-$2 $-$ 10.3&   127.3&  8.2e$-$6\nl
3B 910425&   109&  4&   GRB&  14&    $-$16. $-$ 53.2&   401.6&  4.8e$-$5\nl
3B 910503&   143&  6&   GRB&  27&      0.7 $-$  4.8&   679.9&  7.9e$-$5\nl
3B 910522&   219&456&   BPL&  51&     110. $-$136.&   190.4&  3.5e$-$5\nl
3B 910601&   249&  2&   GRB&  15&      0.0 $-$ 17.9&   306.6&  2.1e$-$5\nl
3B 910619&   394&  1&   GRB&  34&      0.0 $-$ 44.5&   331.6&  4.1e$-$5\nl
3B 910627&   451&  4&   GRB&  16&      0.0 $-$ 12.0&   125.1&  1.5e$-$5\nl
3B 910717&   543&  4&   GRB&  10&      0.0 $-$  6.1&   225.8&  8.3e$-$6\nl
3B 910807&   647&  0&   BPL&  23&      0.0 $-$ 28.3&   162.1&  2.6e$-$5\nl
3B 910814C&   676&  2&   GRB&  18&      0.0 $-$ 54.6&   434.8&  3.0e$-$5\nl
3B 910814&   678&  2&   BPL&  37&      0.0 $-$ 29.4&   444.8&  7.8e$-$5\nl
3B 911031&   973&  3&   GRB&  37&      0.0 $-$ 33.7&   367.2&  3.0e$-$5\nl
3B 911118&  1085&  4&   BPL&  50&      0.0 $-$ 13.7&   161.4&  5.6e$-$5\nl
3B 911126&  1121&  4&   GRB&  28&      19. $-$ 29.9&   251.3&  2.0e$-$5\nl
3B 911127&  1122&  1&   GRB&  36&      0.0 $-$ 28.7&   137.9&  2.4e$-$5\nl
3B 911202&  1141&  7&   GRB&  31&  6.4e$-$2 $-$ 17.9&   354.2&  4.2e$-$5\nl
3B 911209&  1157&  1&   GRB&  18&      0.1 $-$ 23.4&   218.3&  1.8e$-$5\nl
3B 920210&  1385&  5&   BPL&  35&      0.0 $-$ 48.8&   291.5&  4.1e$-$5\nl
3B 920226&  1440&  3&   GRB&  16&      10. $-$ 17.9&   301.6&  1.3e$-$5\nl
3B 920311&  1473&  5&   GRB&  51&       3. $-$ 23.2&   448.2&  8.3e$-$5\nl
3B 920315&  1484&  3&   BPL&   8&      0.0 $-$ 20.1&   174.8&  5.7e$-$6\nl
3B 920406&  1541&  2&   GRB&  14&      67. $-$ 95.2&   223.2&  1.1e$-$4\nl
3B 920513&  1606&  3&   GRB&  28&      12. $-$102.&   206.7&  5.8e$-$5\nl
3B 920525&  1625&  4&   GRB&  35&       4. $-$ 19.7&   447.3&  6.4e$-$5\nl
3B 920622&  1663&  4&   GRB&  60&      0.0 $-$ 24.4&   509.8&  1.0e$-$4\nl
3B 920627&  1676&  S2&   GRB&  29&  6.4e$-$2 $-$ 38.1&   236.7&  2.4e$-$5\nl
3B 920711&  1695&  7&   GRB&  29&      0.0 $-$ 39.2&   348.7&  1.1e$-$4\nl
3B 920718&  1709&  7&   GRB&  13&      0.1 $-$  5.1&   192.2&  9.9e$-$6\nl
3B 920723&  1721&  3&   BPL&  32&  6.4e$-$2 $-$ 30.5&   208.9&  2.6e$-$5\nl
3B 920902&  1886&  5&   GRB&  30&  6.4e$-$2 $-$ 14.6&   472.2&  4.5e$-$5\nl
3B 921003&  1974&  2&   GRB&  21&      0.0 $-$  9.5&    56.5&  1.3e$-$5\nl
3B 921009&  1983&  2&   GRB&  54&  6.4e$-$2 $-$ 29.2&   253.8&  6.3e$-$5\nl
3B 921015&  1989&  4&   BPL&  31&     110. $-$350.&   128.1&  4.4e$-$5\nl
3B 921022&  1997&  2&   GRB&  22&      0.0 $-$ 45.8&    66.1&  2.0e$-$5\nl
3B 921118&  2061&  4&   BPL&  16&      0.0 $-$ 50.6&   250.6&  2.2e$-$5\nl
3B 921123&  2067&  1&   GRB&  44&      12. $-$ 31.7&   243.7&  5.7e$-$5\nl
3B 921207&  2083&  0&   GRB&  36&      0.0 $-$ 14.3&   202.3&  4.9e$-$5\nl
3B 921209&  2090&  1&   GRB&  17&      0.0 $-$ 13.9&   183.0&  1.2e$-$5\nl
3B 921230&  2110& 57&   BPL&  29&      $-$2. $-$ 31.8&   289.7&  2.5e$-$5\nl
3B 930106&  2122&  6&   GRB&  12&  6.4e$-$2 $-$ 73.9&   137.4&  1.9e$-$5\nl
3B 930120&  2138&  0&   GRB&  13&      68. $-$105.&   138.8&  3.7e$-$5\nl
3B 930201&  2156&  1&   BPL&  39&       9. $-$175.&   161.9&  1.2e$-$4\nl
3B 930405&  2286&  6&   GRB&  26&      0.0 $-$ 25.2&   238.1&  2.4e$-$5\nl
3B 930425&  2316&  1&   GRB&  33&      0.0 $-$ 29.4&   173.6&  2.4e$-$5\nl
3B 930506&  2329&  3&   GRB&  26&      0.0 $-$ 13.9&   557.9&  1.2e$-$4\nl
3B 930916&  2533&  3&   GRB&  43&      0.0 $-$ 39.5&   319.8&  7.1e$-$5\nl
3B 930922&  2537&  1&   GRB&  23&  6.4e$-$2 $-$  4.9&    97.0&  1.6e$-$5\nl
3B 931008&  2571&236&   BPL&  43&  3.0e$-$2 $-$164.&    49.1&  5.1e$-$5\nl
3B 931026&  2606&  7&   GRB&  18&       9. $-$ 97.3&   334.6&  3.1e$-$5\nl
3B 931103&  2617&  5&   GRB&  23&  6.4e$-$2 $-$ 18.7&   430.1&  2.5e$-$5\nl
3B 931126&  2661&  1&   GRB&  18&      0.0 $-$ 12.9&   160.8&  2.9e$-$5\nl
3B 931204&  2676&  1&   GRB&  59&      0.0 $-$ 15.8&   336.3&  8.1e$-$5\nl
3B 940206&  2798& 13&   BPL&  65&  3.0e$-$2 $-$ 68.1&   208.9&  1.3e$-$4\nl
3B 940210&  2812&  6&   BPL&  24&      0.0 $-$ 29.1&   176.7&  2.0e$-$5\nl
3B 940217&  2831&  0&   GRB&  32&       7. $-$ 33.7&   790.0&  7.4e$-$5\nl
3B 940228&  2852&  7&   BPL&  25&      0.0 $-$ 38.7&   252.5&  3.6e$-$5\nl
3B 940301&  2855&  4&   GRB&  37&      0.0 $-$ 43.3&   287.1&  6.5e$-$5\nl
3B 940302&  2856&  1&   GRB&  42&      $-$3. $-$153.&   308.8&  2.2e$-$4\nl
3B 940319&  2889&  4&   GRB&  10&  6.4e$-$2 $-$ 63.5&   376.8&  3.6e$-$5\nl
3B 940323&  2891& 45&   BPL&  36&      $-$2. $-$ 24.4&   470.3&  4.3e$-$5\nl
3B 940414&  2929&  4&   GRB&  36&      0.0 $-$ 46.2&   710.0&  4.6e$-$5\nl
3B 940429&  2953&  3&   GRB&  33&      0.0 $-$ 25.1&   165.0&  2.7e$-$5\nl
3B 940526B&  2993&  3&  COMP&   9&      0.0 $-$ 28.8&  3031.0&  2.1e$-$5\nl
3B 940526&  2994&  1&   BPL&  44&       3. $-$ 27.3&   509.5&  5.0e$-$5\nl
3B 940529&  3003&  0&   GRB&  13&      0.0 $-$ 35.6&   279.6&  1.9e$-$5\nl
3B 940619&  3035&  6&   GRB&  13&      0.0 $-$ 62.0&   200.6&  1.9e$-$5\nl
3B 940623&  3042&  1&   GRB&  19&      0.1 $-$ 16.2&   342.1&  1.5e$-$5\nl
3B 940703&  3057&  S5&   BPL&  22&      27. $-$ 92.8&   242.6&  2.3e$-$4\nl
3B 940708&  3067&  6&   GRB&  25&      0.0 $-$  7.7&   316.2&  3.0e$-$5\nl
3B 940810&  3115&  3&   GRB&  21&      11. $-$ 30.7&   254.8&  1.8e$-$5\nl
3B 940817&  3128&  5&   GRB&  43&      20. $-$ 48.0&   372.6&  6.1e$-$5\nl
3B 940826&  3138&  6&   GRB&  18&       8. $-$ 17.4&   142.9&  1.1e$-$5\nl
4B 940921&  3178&  2&   GRB&  22&  3.3e$-$2 $-$ 24.8&   662.0&  5.4e$-$5\nl
4B 941008&  3227&  5&   GRB&  14&      82. $-$ 93.1&   253.0&  1.2e$-$5\nl
4B 941014&  3241&  6&   GRB&  37&      19. $-$ 44.7&   360.6&  3.0e$-$5\nl
4B 941017&  3245&  4&   GRB&  91&       5. $-$ 86.6&   317.5&  1.6e$-$4\nl
4B 941020&  3253&  S5&   GRB&  28&      16. $-$ 70.3&   235.7&  6.7e$-$5\nl
4B 941023&  3255&  4&   BPL&  14&      0.1 $-$ 32.3&   144.9&  2.0e$-$5\nl
4B 941121&  3290&  4&   GRB&  12&      34. $-$ 52.0&   127.0&  1.6e$-$5\nl
4B 941228&  3330&  3&   GRB&  14&      0.1 $-$ 59.6&   404.8&  2.2e$-$5\nl
4B 950104&  3345&  1&   GRB&  17&  3.3e$-$2 $-$ 11.6&   226.8&  1.4e$-$5\nl
4B 950111&  3352&  2&   GRB&  28&  3.3e$-$2 $-$ 42.6&   170.8&  2.6e$-$5\nl
4B 950208&  3408&  6&   GRB&  61&  9.0e$-$2 $-$ 28.3&   410.2&  5.5e$-$5\nl
4B 950211&  3415&  5&   GRB&  25&  2.6e$-$2 $-$ 54.6&   169.7&  2.9e$-$5\nl
4B 950301&  3448&  3&   BPL&  11&       180. $-$338.&   196.1&  2.3e$-$5\nl
4B 950305&  3458&  S4&   GRB&  17&  3.3e$-$2 $-$ 23.0&   224.8&  2.1e$-$5\nl
4B 950325&  3481&  2&   GRB&  21&      37. $-$ 47.6&   280.8&  2.7e$-$5\nl
4B 950401&  3489&  5&   GRB&  15&      0.1 $-$ 18.6&   331.0&  2.6e$-$5\nl
4B 950403&  3491&  3&   GRB&  52&       2. $-$ 17.3&   191.3&  4.4e$-$5\nl
4B 950403B&  3492&  5&   GRB&  26&       3. $-$  9.0&   406.9&  4.3e$-$5\nl
4B 950425&  3523&  6&   BPL&  58&  3.3e$-$2 $-$ 28.9&   357.7&  1.1e$-$4\nl
4B 950513&  3571&  5&   GRB&  22&  9.2e$-$2 $-$ 36.4&   282.4&  2.3e$-$5\nl
4B 950522&  3593&  2&   BPL&  14&  2.6e$-$2 $-$ 18.4&   395.1&  2.1e$-$5\nl
4B 950701&  3657&  4&   GRB&  26&  3.2e$-$2 $-$  9.6&   256.6&  2.3e$-$5\nl
4B 950701B&  3658&  5&   GRB&  30&  2.6e$-$2 $-$ 11.4&   233.8&  2.0e$-$5\nl
4B 950804&  3734&  4&   GRB&  15&      0.1 $-$  3.7&   311.0&  2.0e$-$5\nl
4B 950818&  3765&  1&   GRB&  31&      52. $-$ 73.3&   255.3&  3.4e$-$5\nl
4B 950909&  3788&  3&   GRB&  23&  1.5e$-$8 $-$ 67.6&   183.3&  3.9e$-$5\nl
4B 951011&  3860&  5&   GRB&  16&  3.3e$-$2 $-$ 29.8&   516.7&  2.9e$-$5\nl
4B 951016&  3870&  5&   GRB&  16&  2.6e$-$2 $-$  5.4&    97.4&  1.2e$-$5\nl
4B 951102&  3891&  2&   GRB&  21&        26. $-$ 42.4&   164.2&  1.3e$-$5\nl
4B 951203&  3930&  0&   GRB&  20&  3.2e$-$2 $-$ 21.1&   546.8&  5.4e$-$5\nl
4B 951219&  4039&  6&  COMP&  19&  2.5e$-$2 $-$ 43.0&  1301.8&  3.7e$-$5\nl
4B 960114&  4368&  S0&   BPL&  25&        15. $-$ 27.2&   119.2&  1.2e$-$4\nl
4B 960124&  4556&  5&   GRB&  23&  2.6e$-$2 $-$  5.0&   251.4&  1.6e$-$5\nl
4B 960201&  4701&  1&   GRB&  21&       7. $-$ 26.3&   227.2&  2.1e$-$5\nl
4B 960321&  5299&  S0&   BPL&  46&        41. $-$ 69.4&   107.8&  4.5e$-$5\nl
4B 960322&  5304& 57&   GRB&  59&  2.4e$-$2 $-$ 26.9&   278.3&  7.3e$-$5\nl
4B 960529&  5477&  1&  COMP&  20&  9.1e$-$2 $-$ 10.0& 10423.7&  3.5e$-$5\nl
4B 960605&  5486& 23&   BPL&  46&        60. $-$ 89.4&   220.9&  5.9e$-$5\nl
4B 960607&  5489&  1&   GRB&  26&        30. $-$ 55.0&   280.6&  2.9e$-$5\nl
4B 960623&  5512&  5&   BPL&  13&  3.3e$-$2 $-$ 30.0&   165.7&  1.8e$-$5\nl
4B 960807&  5567&  0&   GRB&  32&  9.0e$-$2 $-$ 17.8&   300.2&  2.7e$-$5\nl
4B 960808&  5568&  6&   GRB&   8&      0.2 $-$  5.6&   513.3&  1.9e$-$5\nl
4B 960831&  5591& 67&   BPL&  49&      $-$9. $-$167.&   282.8&  5.6e$-$5\nl
4B 960924&  5614&  S6&   BPL&  27&       7. $-$ 13.1&   201.2&  1.5e$-$4\nl
4B 961001&  5621&  2&   GRB&  25&  2.6e$-$2 $-$ 10.3&   211.0&  2.9e$-$5\nl
4B 961009&  5629&  6&   BPL&  35&  3.3e$-$2 $-$ 15.9&   189.8&  2.1e$-$5\nl
4B 961029&  5649& 15&   GRB&  59&       4. $-$100.&   239.4&  1.7e$-$4\nl
4B 961102&  5654& 15&   GRB&  63&  3.0e$-$2 $-$100.&   252.2&  6.0e$-$5\nl
4B 961202&  5704&  0&   GRB&  13&  3.0e$-$2 $-$  5.2&   127.0&  1.2e$-$5\nl
4B 970111&  5773&  0&  COMP&  57&  2.9e$-$2 $-$ 21.8&   187.7&  4.4e$-$5\nl
4B 970202&  5995&  1&   BPL&  60&  9.0e$-$2 $-$ 21.6&   265.8&  7.3e$-$5\nl
4B 970223&  6100&  6&   BPL&  35&  9.0e$-$2 $-$ 18.4&   250.9&  4.2e$-$5\nl
4B 970306&  6115&  2&   BPL&  15&      0.1 $-$108.&   304.5&  4.0e$-$5\nl
4B 970315&  6124&  S2&   BPL&  47&  2.6e$-$2 $-$ 19.2&   207.1&  6.4e$-$5\nl
\tablenotetext{a}{Burst names are from Meegan et al. 1996, and continue
with a `4B' prefix for bursts that occurred after the end of the 3B Catalog.}
\tablenotetext{b}{Entries with multiple detectors indicate that summed data
were used for the analysis. Entries prefixed by `S' indicate that SD data were
used instead of missing LAD data.}
\tablenotetext{c}{`GRB': the spectral form of Band et al. (1993);
`BPL': broken power law; and `COMP': power law with a high$-$energy exponential
cut$-$off, in the same $E_{\rm{peak}}$ parametrization as the GRB model.}
\tablenotetext{d}{Fluences as determined over the selected time interval for 
the given spectral form; differs from the 3B Catalog values.}
\enddata
\end{deluxetable}